# Copyright notice

# Experimental Observation of Localized Interfacial Phonon Modes

Zhe Cheng[1,a)], Ruiyang Li[2,a)], Xingxu Yan[3,9,a)], Glenn Jernigan[4], Jingjing Shi[1], Michael E. Liao[5], Nicholas J. Hines[1], Chaitanya A. Gadre[6], Juan Carlos Idrobo[7], Eungkyu Lee[8], Karl D. Hobart[4], Mark S. Goorsky[5], Xiaoqing Pan[3,6,9,*], Tengfei Luo[2,*], Samuel Graham[1,*]

[1] George W. Woodruff School of Mechanical Engineering, Georgia Institute of Technology, Atlanta, GA 30332, United States

[2] Department of Aerospace and Mechanical Engineering, University of Notre Dame, Notre Dame, IN 46556, United States

[3] Department of Materials Science and Engineering, University of California, Irvine, CA 92697, United States

[4] U.S. Naval Research Laboratory, 4555 Overlook Avenue SW, Washington, DC 20375, United States

[5] Materials Science and Engineering, University of California, Los Angeles, Los Angeles, CA 91355, United States

[6] Department of Physics and Astronomy, University of California, Irvine, CA 92617, United States.

[7] Center for Nanophase Materials Sciences, Oak Ridge National Laboratory, Oak Ridge, TN 37831, United States

[8] Department of Electronic Engineering, Kyung Hee University, Yongin-si, Gyeonggi-do 17104, South Korea

[9] Irvine Materials Research Institute, University of California, Irvine, CA 92697, USA.

a) These authors contributed equally.

*Corresponding authors: sgraham@gatech.edu; tluo@nd.edu; xiaoqinp@uci.edu

**Abstract**

Interfaces impede heat flow in micro/nanostructured systems. Conventional theories for interfacial thermal transport were derived based on bulk phonon properties of the materials making up the interface without explicitly considering the atomistic interfacial details, which are found critical to correctly describing thermal boundary conductance (TBC). Recent theoretical studies predicted the existence of localized phonon modes at the interface which can play an important role in understanding interfacial thermal transport. However, experimental validation is still lacking. Through a combination of Raman spectroscopy and high-energy resolution electron energy-loss spectroscopy (EELS) in a scanning transmission electron microscope, we report the first experimental observation of localized interfacial phonon modes at ~12 THz at a high-quality epitaxial Si-Ge interface. These modes are further confirmed using molecular dynamics simulations with a high-fidelity neural network interatomic potential, which also yield TBC agreeing well with that measured from time-domain thermoreflectance (TDTR) experiments. Simulations find that the interfacial phonon modes have obvious contribution to the total TBC. Our findings may significantly contribute to the understanding of interfacial thermal transport physics and have impact on engineering TBC at interfaces in applications such as electronics thermal management and thermoelectric energy conversion.

# Introduction

Interfaces impede heat flow in micro/nanostructured systems, which highlight the growing importance of thermal management of microelectronics and devices for energy conversion and storage.[1,2] Thermal boundary conductance (TBC, *G*), the property characterizing thermal transport across an interface, is defined by a temperature drop ($\Delta T$) across an interface for a given heat flux (*J*) as $G=J/\Delta T$.[3,4] For insulator or semiconductor-related interfaces, lattice vibrations (phonons) are the dominant heat carriers for interfacial thermal transport. At the microscale, thermal transport across an interface is usually described by the phonon gas model as phonons impinging the interface, which results in a portion of the energy passing through the interface as characterized by a transmission coefficient. TBC is thus often described as $G = \left(\sum_j \int v_\omega c_\omega t_\omega d\omega_j\right)/4$, where $v_\omega, c_\omega, t_\omega$ are respectively group velocity, heat capacity, and transmission coefficient of the incident phonon with frequency $\omega$, and polarization *j*.[4] A critical task in the interfacial thermal transport research has been calculating the transmission coefficient.[3,5] The acoustic mismatch model (AMM) and diffuse mismatch model (DMM) are traditional models used for this purpose, and they calculate the transmission coefficient based on the phonon properties of the bulk materials making up the interface.[6-8] However, such treatments are problematic as they ignore any microscopic details of the interface. It has been proven that factors like bond strength, interface roughness, and atomic mixing can significantly impact TBC.[1,9-12] Atomistic Green's Function (AGF) is able to explicitly include realistic atomistic structures in the TBC calculation, which can yield better prediction compared to DMM or AMM.[13] However, phonon transmission coefficients calculated using AGF are still expressed as a function of phonons of the bulk materials.[11,14,15]

In recent years, several molecular dynamics (MD) studies[16-19] have shown that phonon modes that are unique to the interface (i.e., interfacial phonon modes) can play important roles in interfacial thermal transport. These modes are localized in the interfacial region due to the special interatomic bonding environment not seen in bulk materials. However, these MD simulations were either based on empirical interatomic potentials that were not specifically developed for interfaces or employed approximation for interfacial interactions despite the use of first-principles force constants.[16-19] It remains unclear if such interfacial phonon modes indeed exist at real interfaces, which calls for the experimental observation of such modes. The univocal confirmation of the existence of interfacial phonon modes is going to be critical to correctly understanding interfacial thermal transport physics and potentially engineering TBC. However, to date, experiments have been impeded by the technical difficulties related to probing localized modes at relevant size scales and having a properly chosen interface that is free of other factors (e.g., phonon polariton in polar materials, roughness and etc.).[5,20-22]

In this work, we, for the first time, confirm the existence of interfacial phonon modes by creating high-quality Si-Ge interfaces using molecular beam epitaxy (MBE) and detecting the localized modes by Raman spectroscopy and high-energy-resolution electron energy-loss spectroscopy (EELS) in a scanning transmission electron microscope (STEM). Because both Si and Ge are nonpolar materials, there are no strong delocalized phonon polariton modes in the acquired EELS signal which enables the spatial resolution of the characterization to be atomic-scale.[20] We further confirm that the detected localized modes from these experiments are indeed interfacial phonon modes using MD simulations with high-fidelity neural network potential (NNP) trained by first-principles calculations specifically for the Si-Ge interface. The calculated TBC from MD agrees well with the experimentally measured value using time-domain thermoreflectance (TDTR). Finally, spectral analysis in MD shows that the interfacial phonon modes can obviously contribute to the overall TBC, despite its limited population and localized nature.

**Results and Discussions**

To fabricate interfaces that are clean and sharp for interfacial phonon mode detection, we choose the nearly lattice-matched Si-Ge interface and grew high-quality interfaces using MBE which enables a precise control of the sample structure. Two sample architectures were grown, each fabricated for interfacial phonon mode detection and TDTR measurement, respectively. The detailed growth processes can be found in Supplementary Information (SI). Sample 1 consists of a 10-nm Si cap on a 300-nm Ge layer grown on a Si substrate and it is used for Raman (Fig. 1a) and EELS measurements (Fig. 2a). Measurement details are included in the SI. Such a structure of Sample 1 is chosen since the absorption depth of Si is large (774 nm), tens of times greater than that of Ge, at the Raman laser wavelength (488 nm),[23] and thus the 10-nm Si cap layer can allow the Raman laser to transmit through to reach the Si-Ge interfaces. Sample 2 is a 250-nm Ge layer grown on a Si substrate and is used for TBC measurements, where the Ge layer thickness is optimized to maximize the sensitivity of the TBC of the Si-Ge interface in TDTR measurement. To confirm the quality of the Si-Ge interface, high-angle annular dark-field scanning transmission electron microscopy (HAADF-STEM) imaging technique is used to characterize the interface of Sample 1. In Fig. 1a, the Z-contrast HAADF-STEM image shows that the interface is very sharp despite an atomic mixing layer with a thickness of about one unit cell (~0.8 nm).

As a comparison to the Raman spectrum of Sample 1 which involves a Si-Ge interface, we have also measured the Raman spectra of a pure Ge wafer and a pure Si wafer. Figure 1b shows that the Raman peak of bulk Ge is at ~9 THz and that of Si is at ~15.6 THz. The Si peak in Sample 1 is redshifted by 0.15 THz compared to pure Si because the epitaxial Si cap layer is slightly stretched on the Ge wafer to accommodate the ~4% lattice mismatch. Other than these two peaks originated from Ge and Si, an additional peak around 11.3-12.2 THz shows up in Sample 1, which may be attributed to the Si-Ge interface.

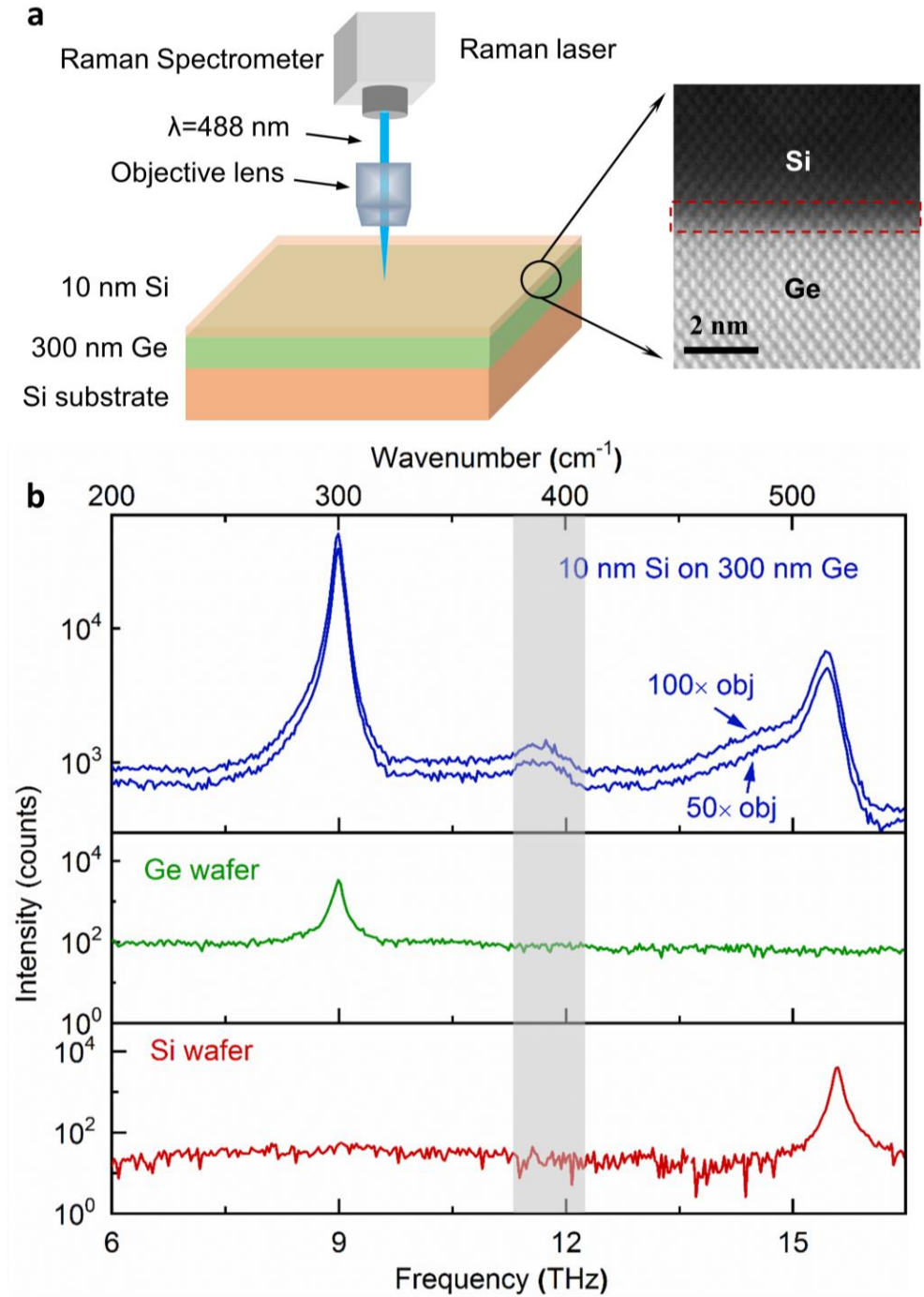

**Figure 1. Raman detection of interfacial modes at the Si-Ge interface.** (a) Schematic diagram of Raman measurements on Sample 1 and its HAADF-STEM image showing a high-quality Si-Ge interface. (b) Raman spectra of Sample 1, a Ge wafer, and a Si wafer. Two objectives (50× and 100×) were used to perform the Raman measurements on Sample 1 and the measurement details

can be found in SI. A distinct Raman peak around 11.3-12.2 THz from Sample 1 is not seen in bulk Ge or Si wafers, which is attributed to the interfacial modes.

To further confirm that this peak originates from the interface, we employ high-energy-resolution (routinely ~1.7-1.9 THz with an exposure time of 1 s) EELS, in a STEM, with a probe size of 1.5 Å to spatially resolve the phonon signal around the Si-Ge interface (see details in SI). Figure 2a shows the schematic diagram of the EELS measurements. The peaks in the EELS signal are the energies of vibrational phonon modes in the sample that inelastically interact with the incident electron probe. Figure 2b shows a line-scan of the vibrational spectra across the interface as illustrated by the arrows in Fig. 2a. Under our experimental setup with a large convergence semi-angle (33 mrad), the as-acquired vibrational spectra contain the momentum-integrated vibrational modes throughout the entire Brillouin zone and are comparable with the phonon density of states (PDOS).[24] The dashed rectangle highlights the region around the Si-Ge interface, which clearly shows the existence of localized vibration modes unique to the interfacial region.

Three representative EELS vibrational spectra from Si (red), Ge (green) and the interface (blue) taken at the locations marked in Fig. 2a are shown in Fig. 2c. The vibrational spectrum of the interface region includes the peaks from Si, Ge and the interfacial vibration modes. A linear regression fitting of the EELS signal at the interface, as shown in Supplementary Fig. 4, confirms that the intensity peak near the interface is not due to the overlapping of the Si and Ge signals. Figure 2d shows the intensities of vibrational spectra at 11.6, 12.0, and 12.4 THz as a function of distance to the interface which are marked by dashed lines in Fig. 2c. Clear peaks of the vibrational spectral intensity are observed around the interface at 11.6 and 12.0 THz. The fact that these peaks are unique to the interface and have a frequency of 11.2-12.3 THz – a frequency also observed in Raman spectra, indicate that this is a localized interfacial phonon mode. Leveraging the high spatial resolution (1.5 Å) capability of STEM-EELS, this interfacial vibrational mode is determined to be confined to within ~1.2 nm of the interface. By combining the Raman and high-energy-resolution EELS-STEM results above, we can claim that detected modes around 11.2-12.3 THz are indeed interfacial phonon modes of the Si-Ge interface and they are localized at the interface.

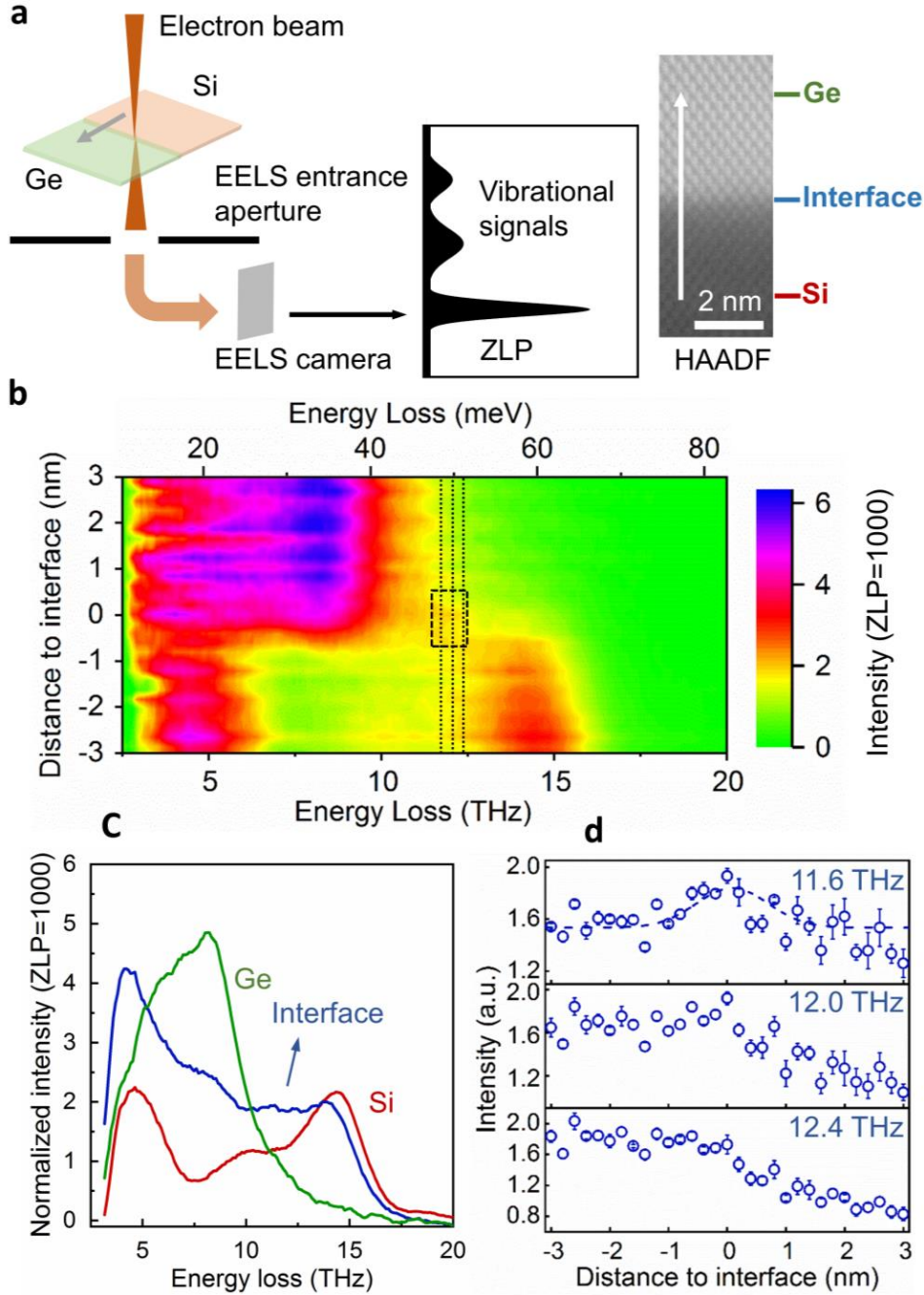

**Figure 2. EELS-STEM spatially resolves the interfacial phonon modes.** (a) Schematic diagram of EELS measurement of vibrational modes around the Si-Ge interface, and the path of a line scan. (b) The line profile of the vibrational spectra across the interface. The dash line rectangle labels the interfacial vibration modes. (c) Three representative vibrational spectra with a total acquisition time of 200 s from Si (red), Ge (green) and interface (blue) detected from EELS, as marked in (a).

(d) The integrated vibrational signal at 11.6, 12.0, and 12.4 THz in a window with width of 0.48 THz. The measured signal at 11.6 THz overlaps with a Gaussian fitting curve with a full width half maximum (FWHM) of 1.2 nm (less than two unit cells).

To further understand the interfacial modes, we use MD simulations to model the Si-Ge interface with the level of interfacial mixing (~0.7 nm thick) similar to those observed in the HAADF STEM image from experimental measurements (see Fig. 3a and SI for simulation details). In order to reproduce the interfacial modes, a high-fidelity NNP for the Si-Ge interface is specifically developed by training against first-principles density functional theory (DFT) calculations (see SI for details). The NNP-MD scheme has been proved capable of simulating semiconductor materials and predicting thermal properties with accuracy comparable to first-principles calculations.[25,26] Using MD simulations with this NNP, PDOS at different locations throughout the simulation domain along the direction perpendicular to the Si-Ge interface is calculated (see SI for details).

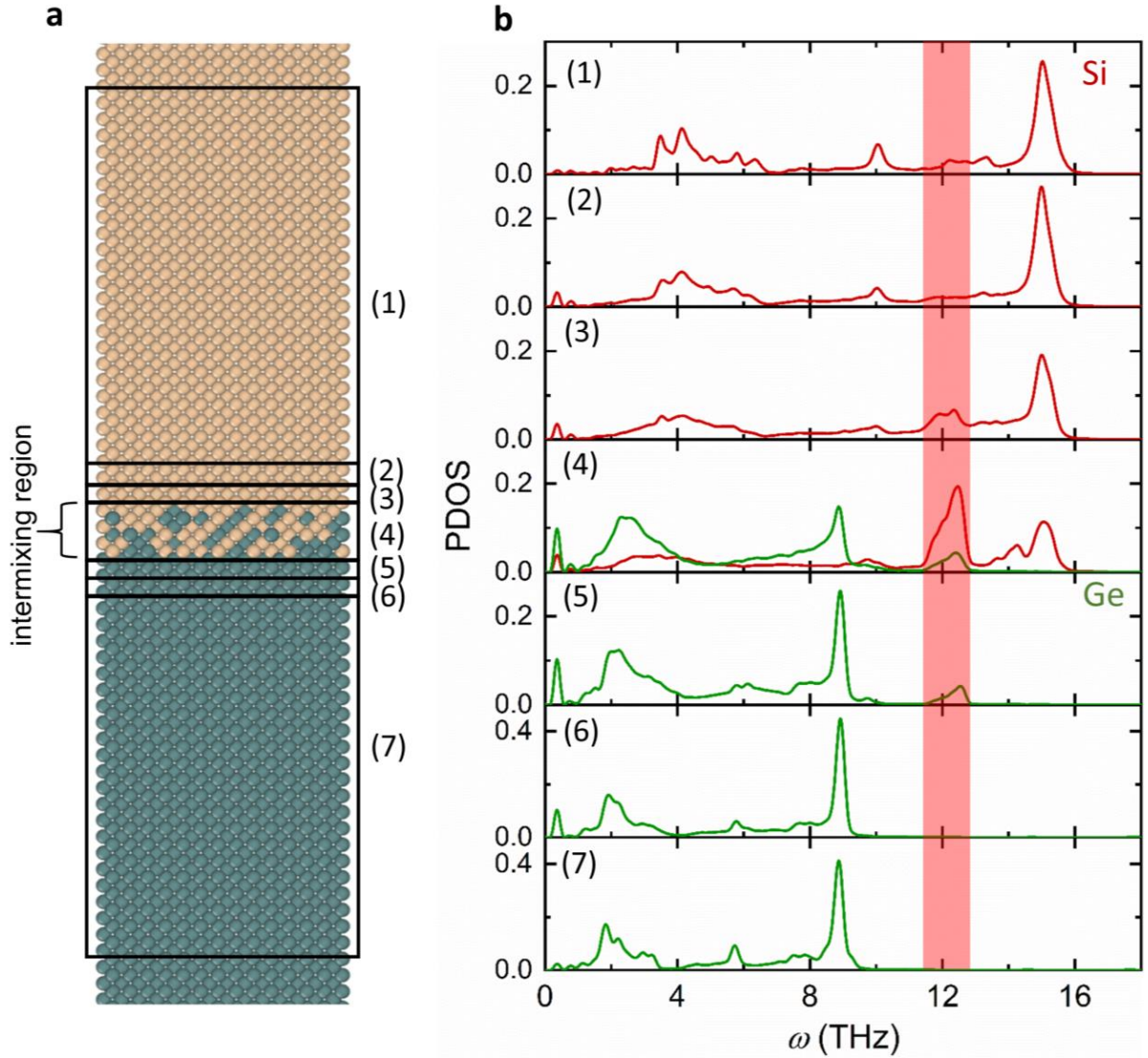

**Figure 3. MD simulation confirming interfacial modes.** (a) Schematic diagram of the MD-simulated system and the regions where PDOS are calculated. (b) PDOS by MD simulations at different locations throughout the domain along the direction perpendicular to the Si-Ge interface. Each panel is the PDOS of the atoms at a location at or away from the interface (middle of the interface mixture) indicated in (a). Interfacial modes exist at the intermixing region and the first adjacent atomic layer on each side.

As shown in Fig. 3b, each panel is the PDOS of the atoms at a certain location shown in Fig. 3a. The interfacial region is composed of the mixed region (~0.7 nm thick) and the first layer away from the mixture on each side. For the regions which are 10 nm away from the interface, the PDOS of both Si and Ge are converged to the ones of bulk and are not affected by the interface. The PDOS peaks of Si and Ge optical phonons are ~15 THz and ~9 THz, which are consistent with the Raman peaks. Near the interface, a peak at ~11.4-12.8 THz emerges for both the Si and Ge atoms, which corresponds well with the interfacial modes detected from Raman and EELS. This calculated interfacial mode peak agrees better with our experiments compared to other MD simulations results using empirical potentials. Chalopin *et al*. obtained such a peak at ~13.5 THz using a Stinger-Weber potential,[16] while Gordiz *et al*. found the peak between 12-13 THz using a Tersoff potential.[17] Our calculated peak is similar to that calculated from first-principles force constants (~12 THz), which however used Si force constants to approximate those for Ge and interfacial interactions.[18] We note that all these previous studies modeled sharp interfaces without atomic mixing effects considered. We have also performed simulation on a sharp Si-Ge interface and also found a peak around 11.6-12.7 THz (Supplementary Fig. 8, see SI).

The peak is very dominant in the intermixed region, but it is also spatially extended to the first layers next to the intermixing layer in both Ge and Si sides. The delocalization length is around 1.1 nm in the interfacial region, which agrees very well with the EELS detected interfacial mode span of ~1.2 nm around the interface (Fig. 2d). Supplementary Fig. 7 shows the eigenvectors for an interfacial mode which delocalize into both Ge and Si sides. A prior work using the nearest neighbor in the first-principles force cutoff was not able to capture this delocalization effect.[18]

To understand how these interfacial modes contribute to interfacial transport, we performed non-equilibrium MD to calculate the TBC (see SI for details). The calculated TBC is ~250 MW m$^{-2}$ K$^{-1}$ at 300 K. We further use spectral analysis to calculate the contribution of phonons with different frequencies to the TBC (see SI for details). This does not only allow us to directly visualize the role of interfacial modes, but also allows quantum correction to be applied so that the MD-calculated TBC can be directly compared to our experimental TDTR measurement result. As can be seen, there is an obvious peak of the frequency-dependent TBC around the interfacial phonon mode region (11.4-12.8 THz) from our spectral analysis (Fig. 4a). Since MD are classical simulations, all phonons are equally excited. However, in reality, phonon follows the quantum Bose-Einstein distribution. Such an error can be corrected by applying a quantum correction by weighing the contribution of different frequencies with the ratio of the quantum and classical heat capacities of that frequency (see SI for details). The quantum correction leads to reduced contributions from higher frequency modes as they are not fully excited, but the contribution from the interfacial modes is still obvious. We integrate the frequency-dependent TBC to obtain the cumulative TBC, and find that the interfacial modes contribute ~5% of the total TBC (Fig. 4b). The quantum-corrected TBC (231 MW m$^{-2}$ K$^{-1}$) agrees very well with our TDTR measurement (244 -66/+99 MW m$^{-2}$ K$^{-1}$). We note that these calculations are on the mixed interface. For a sharp interface, the calculated TBC is 197 MW m$^{-2}$ K$^{-1}$, which is lower than the mixed interface and farther away from the measured TBC. This is consistent with previous findings that interface mixing can enhance TBC.[11,12] Figure 4c also show the comparison of TBC values of Si-Ge interface calculated from DMM, AGF (both harmonic and anharmonic) and previous MD simulations using different potentials. The NNP-MD results for both the mixed interface and the sharp interface respectively agree well with their counterparts from AGF calculations, highlighting the important of correctly capturing the microscopic interface conditions for TBC prediction. On the other hand, other models (e.g., DMM) that ignore such microscopic interfacial details show inferior agreement with experimental measured TBC. The calculation details of non-equilibrium Landauer approach based on DMM can be found in the SI. The fact that NNP-MD is accurate in predicting TBC also give confidence on its predicted non-trivial contribution of interfacial modes to the TBC.

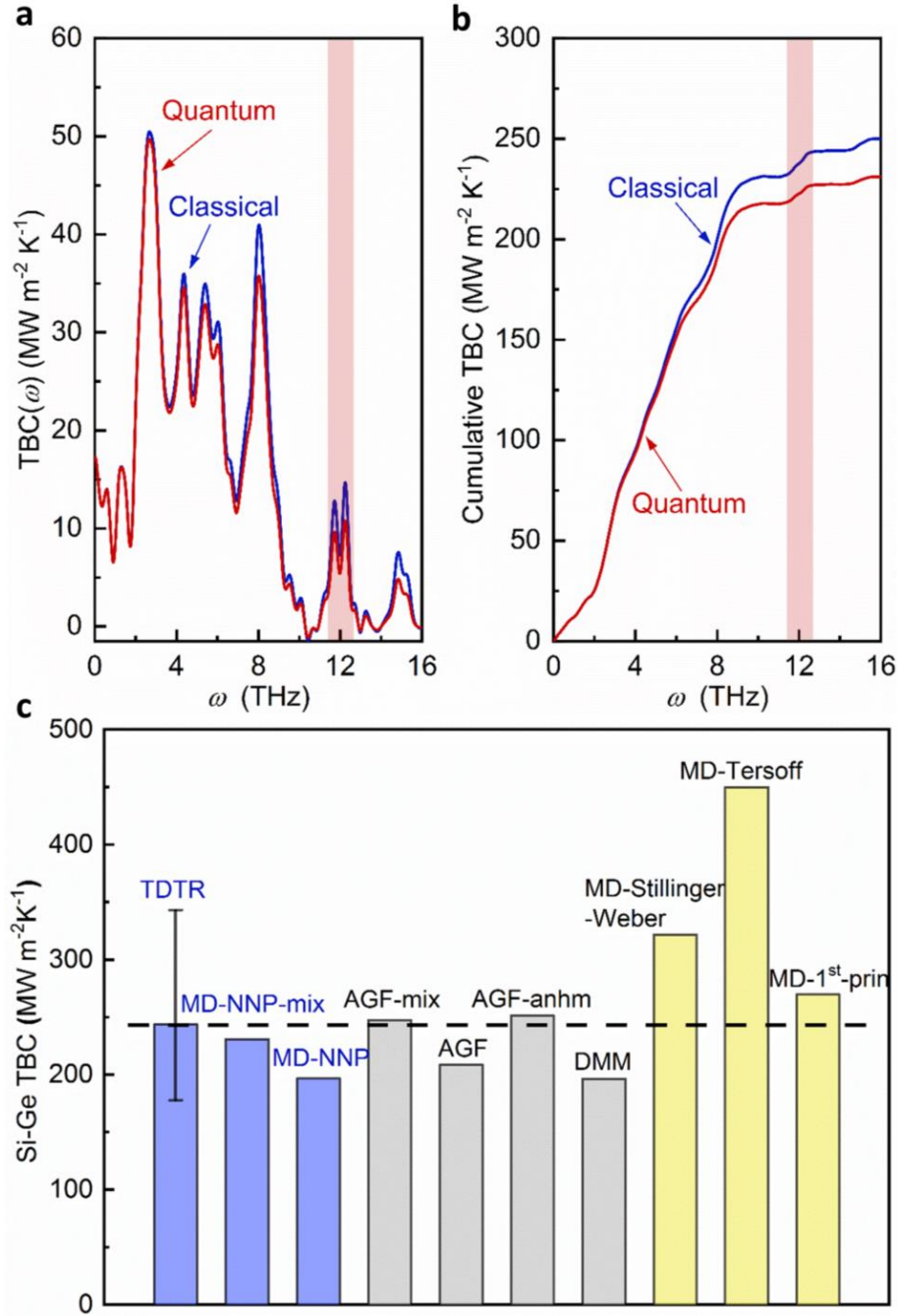

**Figure 4. TBC of Si-Ge interfaces at room temperature**. (a) Spectral TBC of the Si-Ge interface. (b) Cumulative TBC of the Si-Ge interface. (c) Comparison of the measured TBC with calculated TBC values. The "MD-NNP-mix" and "MD-NNP" are TBC values of mixed and ideal Si-Ge interfaces by MD with NNP. "AGF" and "AGF_mix" are TBC values of an ideal Si-Ge interface and a Si-Ge interface with 6-atomic-layer mixing calculated by atomistic Green's function

(AGF).[11] "AGF_anhm" is the TBC of a perfect Si-Ge interface calculated by AGF which includes both harmonic and anharmonic contributions.[13] "DMM" is the TBC calculated by the non-equilibrium Landauer approach which considers the non-equilibrium effect at the interface and uses DMM to calculate transmission. The TBC of perfect Si-Ge interfaces calculated with other interatomic potentials such as Stillinger-Weber potential, Tersoff potential, and first-principles force constant potential are also included for comparison.[18,19,27,28]

In summary, for the first time, we experimentally observed the existence of localized interfacial vibrational modes within 1.2 nm at an epitaxial Si-Ge heterointerface by Raman spectra and high-energy-resolution EELS in a STEM. These interfacial modes are found to be reproducible using MD simulations with high-fidelity NNP, which also yield TBC agreeing favorably to TDTR measurements. The spectral analysis in MD indicates that these interfacial modes, also localized, can have non-trivial contribution to TBC. This work paves the way to fundamentally understanding heat transport across realistic interfaces and may stimulate new theories of interfacial thermal transport. It will impact applications of electronics thermal management and thermoelectric energy conversion.

## Methods

Detailed description of materials growth, Raman measurements, TDTR measurements, XRD measurements, STEM and EELS measurements, non-equilibrium Landauer approach, neural network potential, and non-equilibrium MD simulations are provided in the Supplementary Information.

## Acknowledgements

We would like to acknowledge the financial support from Office of Naval Research MURI Grant No. N00014-18-1-2429. We thank helpful discussions with Asegun Henry, Andrew Rohskopf, and Tianli Feng. The TEM work was supported by the Department of Energy (DOE), Office of Basic Energy Sciences, Division of Materials Sciences and Engineering under Grant DE-SC0014430, and partially supported by the Center for Nanophase Materials Sciences, which is a DOE Office of Science User Facility (JCI). The authors acknowledge the use of facilities and instrumentation at the UC Irvine Materials Research Institute (IMRI), which is supported in part by the National Science Foundation through the UC Irvine Materials Research Science and Engineering Center (DMR-2011967).

## Author contributions

## Competing interests

The authors declare no competing interest.

## Data availability

The data that support the findings of this study are available from the corresponding authors upon reasonable request.